# EOMM: An Engagement Optimized Matchmaking Framework

Zhengxing Chen [*]
Northeastern University
czxttkl@gmail.com

Su Xue
Electronic Arts, Inc.
sxue@ea.com

John Kolen
Electronic Arts, Inc.
jkolen@ea.com

Navid Aghdaie
Electronic Arts, Inc.
naghdaie@ea.com

Kazi A. Zaman
Electronic Arts, Inc.
kzaman@ea.com

Yizhou Sun
University of California, Los Angeles
yzsun@cs.ucla.edu

Magy Seif El-Nasr
Northeastern University
m.seifel-nasr@neu.edu

## ABSTRACT

Matchmaking connects multiple players to participate in online player-versus-player games. Current matchmaking systems depend on a single core strategy: create *fair games* at all times. These systems pair similarly skilled players on the assumption that a fair game is best player experience. We will demonstrate, however, that this intuitive assumption sometimes fails and that matchmaking based on fairness is not optimal for engagement.

In this paper, we propose an Engagement Optimized Matchmaking (EOMM) framework that maximizes overall player engagement. We prove that equal-skill based matchmaking is a special case of EOMM on a highly simplified assumption that rarely holds in reality. Our simulation on real data from a popular game made by Electronic Arts,Inc. (EA) supports our theoretical results, showing significant improvement in enhancing player engagement compared to existing matchmaking methods.

## 1. INTRODUCTION

Player-versus-Player (PvP) is a mode of video game in which multiple players directly engage in competition or combat. PvP games, which cover many popular genres, such as multiplayer online battle arena (MOBA), first-person shooting (FPS), and e-Sports, have increased worldwide popularity in recent years. For example, *League of Legends*, one of the most played MOBA games, has 90 million summoner names registered, 27 million unique daily players and 7.5 million concurrent users [30, 41]. As data released by [39] shows, e-Sports is estimated to have 188 million viewership and 748 million dollar worth market in 2015 and the numbers are expected to grow continuously.

Matchmaking is the process that connects players to form PvP matches. In practice, a matchmaking system takes practical limitations, such as players' geo-location and network latency, into consideration. For example, cross-ocean pairing is not good for player experience. Beyond technical constraints, the strategy various matchmaking systems employ is *creating fair games*. This strategy relies on the assumption that matching closely skilled players tend to create competitive games which are desired by players [21]. In order to establish player skills, numerous models have been studied, such as Elo [16], Glicko [20] and TrueSkill [23].

Are fairly matched games always beneficial for player experience? This fundamental, yet intuitive, assumption is worthy of deep investigation. We can challenge it with a few examples. Consider a cautious player who cares about protecting his rank among friends, and a risk taker who enjoys difficult matches. Pairing them with the similarly skilled opponents will affect these players very differently. Even for the same player, their expectation on the coming match when they just lost three games in a row can be very different from that when they recently performed well. In Table 1, we show an example that churn risks vary drastically upon players' recent match outcomes in a popular PvP game made by EA. These facts lead to two key insights: (1) the effectiveness of matchmaking needs to be measured quantitatively; and (2) matchmaking should depend on dynamic and individual *player states*.

In this paper, we propose a new matchmaking framework, Engagement Optimized Matchmaking (EOMM). By formulating matchmaking into an optimization problem, we pair players in order to maximize the overall player engagement, or equivalently, minimize the overall player disengagement. First, we measure a player's disengagement by their churn risk after each matchmaking decision. Here *churn* refers to no gameplay within a period of time, such as a week. Second, we model all players who wait in the matchmaking pool as a complete graph, where each player is a node, and an edge between two players is their sum churn risks if paired. The churn risk depends on individual player states at the moment of matchmaking. Last, we can achieve engagement optimized matchmaking efficiently by solving a *minimum weight perfect matching* (MWPM) problem that finds non-overlapping pairs with the minimal sum of edge weights on a complete graph.

---
[*] This work was done when Zhengxing Chen was an intern student at Electronic Arts,Inc.

**Table 1: An example about the impact of player states on their engagement.** Data is from a popular PvP game made by EA. Average churn risks vary drastically upon players' recent three match outcomes (*(W)in*, *(L)ose* or *(D)raw*). Churn risk is measured by the ratio of the players who stop playing within a period time (7 days in this table) after a match. The churn risk of some states with repeated losses (5.1%) is almost twice as much as those of other "safer" states (2.6%-2.7%).

| Last 3 Outcomes | Churn Risk |
|---|---|
| DLW \| LLW \| LDW \| DDD | 2.6% - 2.7% |
| ... | ... |
| WWW | 3.7% |
| ... | ... |
| DLL \| LWL \| LDL | 4.6% - 4.7% |
| WWL | 4.9% |
| LLL | 5.1% |

EOMM provides a solid theoretical framework for matchmaking analysis. With it, we prove that equal-skill based matchmaking is a special case of EOMM on a simplified and often inapplicable assumption about player states. The generic EOMM instead proves to be optimal across a wide range of contexts.

For system development, EOMM is both flexible and computationally feasible. The optimization objective can be tuned for various interests, e.g., in-game time, or even spending. Furthermore, EOMM consists of three components: a skill model, a churn prediction model and a graph matching model. All can be efficiently implemented and independently upgraded. We built a simulated system based on real data of a popular game made by Electronic Arts, Inc. (EA), showing significant improvement in enhancing player engagement by EOMM against equal-skill based and other matchmaking methods.

In sum, this paper contains the following contributions. First, we propose an engagement optimized matchmaking framework, i.e., EOMM, which solves matchmaking as an optimization problem of maximizing the overall player engagement. Second, we provide theoretical analysis about the optimality of EOMM and the conditions of the applicability of existing matchmaking methods. Last, we build a simulated system using real game data to show significant advantages of EOMM in retaining players over the existing matchmaking methods.

The rest of this paper is organized as follows. After reviewing the related work, we will present the formulation of matchmaking as an optimization problem on a graph. Then we describe theoretical findings comparing EOMM and other matchmaking methods. We then show the case study applying EOMM on real data. Finally, we will conclude with a discussion of the results and future directions.

## 2. PREVIOUS WORK

### 2.1 Skill Modeling

The motivation behind skill rating is to rank players and to enable skill-based matchmaking. Dating back to 1952, the *Bradley-Terry model* [7] was developed to deal with repeated pairwise comparisons among a group of subjects. In the Bradley-Terry model, a player $i$ is assumed to have a fixed, positive skill scalar, $r_i$, and the winning probability of player $i$ against player $j$ is the ratio of player $i$'s skill in the sum of skills of both players. In its original form, the Bradley-Terry model estimates player skills only after observing all pairwise comparisons. While feasible for small groups of players, requiring $O(n^2)$ matches is prohibitive for large player pools. One can show that the Bradley-Terry model is equivalent to a logistic regression model [2] in which each coefficient $w_i$ corresponds to $\log(r_i)$.

The *Elo system* [16] addresses the relative skill ratings in player-versus-player games, such as chess, with a probabilistic model. Elo captures player performance, $p_i$, as a random variable following a one dimension Gaussian distribution with a mean, $r_i$, and a fixed variance, $\beta^2$, shared by all players. In the Elo system, $r_i$ gets updated depending on the extent of agreement between expected outcomes and real outcomes. For example, a low skill player beating a high skill player yields a large update in adjusting their skill means closer. Unlike the original Bradley-Terry model, $r_i$ can be updated at an ongoing basis, i.e., as soon as after every match of player $i$.

The *Glicko system* [20], a Bayesian ranking rating system, was later introduced. Besides mean player skill, $r_i$, it also models the belief about a player's skill as $RD_i$ (rating deviation). As they play an increasingly number of games, the belief about their skills become stronger hence $RD_i$ decreases. However, $RD_i$ increases when a player ceases to play for long time. To achieve high efficiency, Glicko uses an approximation Bayesian algorithm to update $r_i$ and $RD_i$.

Neither the Bradley-Terry model, the Elo system or the Glicko system was initially applicable to team-oriented games until works such as [23, 24, 29] to generalize these models. For example, *TrueSkill system* [23] extends the Elo system to games with flexible numbers of players and teams.

Researchers have proposed more advanced skill models to capture player skills in multiple facets. The works in [8, 38] model player skills in multi-dimensions such as offensive and defensive abilities. Delalleau et al. [10] proposed a neural network based skill model which learns latent skill embeddings of players and is claimed to outperform TrueSkill in a team based game. There are skill models proposed for specific game genres, such as [11] for chess, [9, 40] for MOBA games and for [3] RTS games.

In our paper, we will compare EOMM with skill based matchmaking methods which leverage skill models. Skill models can also facilitate EOMM in the decision of player assignment.

### 2.2 Matchmaking Strategies

There has been much research ensuring physical criteria of matchmaking services such as network connection quality [1, 27, 28]. Besides physical criteria, matchmaking can be seen as a player modeling technique [45] that extracts player information and delivers adaptive gaming experience [33]. A fair amount of matchmaking systems assume that skill balanced games are good for engagement [21] and hence resort to skill rating algorithms to identifying similarly skilled opponents. Myślak and Deja [32] suggests additional information about player preferences in in-game avatar roles can further improve fairness-based matchmaking systems. A few researchers have explored methods to improve player engagement through matchmaking. Delalleau et al. [10] proposed to train a neural network based architecture which predicts player enjoyment based on their historical statistics. They measured enjoyment by directly asking players for feedback after each match. Whether or not this method can effectively collect sufficient feedbacks has not been demonstrated with real data. Jiménez-Rodrıguez et al. [25] proposed that matchmaking could be based on preferred roles by players. They argue that a fun match should have players act in roles with perceivably joyful role distribution. However, it is still a conceptual, heuristic-based method without experiment showing that such matchmaking system indeed improves concrete engagement metrics. To our best knowledge, we have not seen any exist-ing matchmaking method that formally treats matchmaking as an optimization problem to maximize player engagement.

### 2.3 Player Engagement Prediction

Player engagement can be seen as an objective measurement of user experience in games [6]. Player engagement can be embodied by many specific metrics, such as time or money spent in the game, the number of matches played within a time window, or churn risk. We define churn risk as the proportion of total players stopping playing the game over a period of time.

Churn prediction has been applied within various disciplines for decades, such as telecommunications [17], online advertisements [46] and insurance [31]. Video games have also sparked a number of churn analysis studies. For instance, Weber et al. [43] built a regression model to predict the number of games played. They also used the model to aid game design by identifying the most influential features on player retention. Hadiji et al. [22] established the fundamental in churn prediction in free-to-play (F2P) games by suggesting definitions of various churn behaviors, proposing universal behavioral telemetry, and comparing different machine learning models across five commercial F2P games. Runge et al. [37] not only trained a churn prediction model for a casual social game but also showed how the game can leverage the model to increase the effectiveness of promotions to players. In EOMM, we employ churn prediction as an important building component in engagement optimization.

### 2.4 Graph Matching

In a graph $\mathcal{G} = (V, E)$, a *matching* is a set of pairwise non-adjacent edges [44]; that is, no two edges share a common vertex. A *perfect matching* is a matching with every vertex in $G$ incident on exactly one edge in the matching. In a weighted graph $G$, a *minimum weight matching* (MWM) is the matching with the lowest sum of edge weights. A *minimum weight perfect matching* (MWPM) is the perfect matching with the lowest sum of edge weights.

As will be shown in Section 3, the EOMM framework converts the problem of determining optimal match assignment to the problem of seeking MWPM on a weighted graph. MWM/MWPM have broad applications in other fields, including creating pairs following specific rules in chess tournaments [34], schdeduling training sessions among NASA shuttle cockpit simulators [4] and transmitting images over networks [36]. In a similar spirit, Ólafsson [34] leverages MWPM algorithm to determine opponents. Their goal, however, was to create matches maximally adhering specific rules of chess tournament, which is different than ours to optimize for player engagement.

The first attempt to solve MWPM is the polynomial time *blossom* algorithm proposed by Edmond [14, 15] in 1965. Since then, researchers have steadily improved upon this algorithm. We will compare and discuss those improved methods later when we introduce EOMM.

## 3. ENGAGEMENT OPTIMIZED MATCHMAKING

In this section, we will introduce the EOMM framework which formulates matchmaking as an optimization problem. In contrast with the existing matchmaking methods that heuristically pair similarly skilled co-players, EOMM aims to match players in an optimal way that maximizes overall player engagement. Here we will describe the details of match assignment for 1-vs-1 games. We will discuss how EOMM can be extended to the matches with more players in the final section.

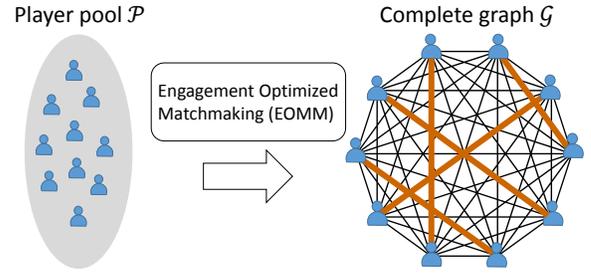

**Figure 1: Model matchmaking on a complete graph. Each node represents a player, and every edge is associated with the sum engagement metric of two players if paired. EOMM amounts to finding an optimal pair assignment on $\mathcal{G}$.**

### 3.1 Optimization Objective

In practice, matchmaking is applied to a pool of players, $\mathcal{P} = \{p_1, \cdots, p_N\}$, who are waiting to start 1-vs-1 matches. We assume $N$ to be an even number such that all players can be paired. The objective of EOMM is to maximize the overall player engagement, or equivalently, minimize the overall player disengagement. We use *churn risk* as a concrete metric of disengagement. The term "churn" is used by convention, which actually represents a status of disengagement, i.e., a player not playing any games within a subsequent time frame, not necessarily a permanent churn. We model the churn risk, $c_{i,j}$, of player $p_i$ after matchmaking with player $p_j$ as a function of both players' states, $c_{i,j} = \Pr(p_i \text{ churns} | \boldsymbol{s_i}, \boldsymbol{s_j}) = c(\boldsymbol{s_i}, \boldsymbol{s_j})$. A player state is a collection of features that profile an individual player, including but not limited to install date, skill, play frequency, performance and etc. We will elaborate on learning $c_{i,j}$ in the subsequent sections. Note that $c_{i,j} \neq c_{j,i}$ since two players in a paired match may be impacted differently. We use a list of player tuples, $\mathcal{M} = \{(p_i, p_j)\}$, to denote a matchmaking result, i.e., a *pair assignment*, in which all players in $\mathcal{P}$ are paired once and only once. Defining the overall player disengagement as the sum of individual churn risks, EOMM seeks for an optimal pair assignment $\mathcal{M}^*$ such that:

$$\mathcal{M}^* = \arg\min_{\mathcal{M}} \sum_{(p_i, p_j) \in \mathcal{M}} c(\boldsymbol{s_i}, \boldsymbol{s_j}) + c(\boldsymbol{s_j}, \boldsymbol{s_i}) \quad (1)$$

We construct a graph, $\mathcal{G}$, to model this environment (see Figure 1). Each player $p_i$ is a node of the graph, who has a player state, $\boldsymbol{s_i}$, before matchmaking. The edge between two players $p_i$ and $p_j$ is associated with a weight $c_{i,j} + c_{j,i}$, which is the expected sum disengagement metric if they are paired. Note that $\mathcal{G}$ is a complete graph in that all pairs of players can be possibly connected. Once all $c_{i,j}$ are computed, finding $\mathcal{M}^*$ in Eqn. 1 is converted to a *minimum weight perfect matching* problem, i.e., finding a pair assignment with the minimal sum weights of edges on graph $\mathcal{G}$.

### 3.2 Predicting Churn Risks

We learn the function $c_{i,j} = c(\boldsymbol{s_i}, \boldsymbol{s_j})$ as a churn prediction problem. In its original form, the churn risk $c_{i,j}$ of player $p_i$ after matchmaking depends on the states from both the player and their opponent. Unfortunately, the well-established churn prediction studies cannot be employed because normally they only use features of players themselves without considering those of opponents. Also, naively feeding both player states as input will double the feature dimension, which makes the prediction unintelligible and harder since much more training data is needed.

One way to simplify the prediction of $c_{i,j}$ is to base it only on player $p_i$'s own state, $\boldsymbol{s}_i$, and the resulting match outcome, $o_{i,j}$, from the view of $p_i$. This works because the opponent's state, $\boldsymbol{s}_j$, such as skill, play history and style, does not directly interact with player $p_i$'s churn risk $c_{i,j}$. It, however, influences the upcoming match outcome, which is directly perceivable by player $p_i$ and thus affects $p_i$'s churn. Once the match outcome $o_{i,j}$ is known, $c_{i,j}$ becomes conditionally independent to the opponent's state, $\boldsymbol{s}_j$. Formally, this property is represented as:

$$c(\boldsymbol{s}_i, \boldsymbol{s}_j, o_{i,j}) = c(\boldsymbol{s}_i, o_{i,j}), \qquad (2)$$

In this paper, we assume that game outcomes are sampled from a finite set, $\mathcal{O}$, such as *Win*, *Lose* and *Draw*. For example, $o_{i,j} = W$ means that $p_i$ wins over $p_j$, while $o_{j,i} = L$ represents the same outcome from the view of $p_j$. We employ standard skill models [16, 20] that use both players' skills to approximately predict game outcome probabilities. We denote player $p_i$'s skill representation as $\boldsymbol{\mu}_i$, which is, for example, Elo score [16] or Glicko mean and RD [20]. Note that $\boldsymbol{\mu}_i$ is part of player state $\boldsymbol{s}_i$. As a result, we have:

$$\Pr(o_{i,j} | \boldsymbol{s}_i, \boldsymbol{s}_j) \approx \Pr(o_{i,j} | \boldsymbol{\mu}_i, \boldsymbol{\mu}_j), \qquad (3)$$

Putting them together, we can efficiently predict the churn risks of paired players in Eqn. 1:

$$c(\boldsymbol{s}_i, \boldsymbol{s}_j) + c(\boldsymbol{s}_j, \boldsymbol{s}_i) \qquad (4)$$

$$= \sum_{o_{i,j} \in \mathcal{O}} \Pr(o_{i,j} | \boldsymbol{s}_i, \boldsymbol{s}_j) \left( c(\boldsymbol{s}_i, \boldsymbol{s}_j, o_{i,j}) + c(\boldsymbol{s}_j, \boldsymbol{s}_i, o_{j,i}) \right) \qquad (5)$$

$$\approx \sum_{o_{i,j} \in \mathcal{O}} \Pr(o_{i,j} | \boldsymbol{\mu}_i, \boldsymbol{\mu}_j) \left( c(\boldsymbol{s}_i, o_{i,j}) + c(\boldsymbol{s}_j, o_{j,i}) \right), \qquad (6)$$

where the first equality is a marginalization on game outcome, $o_{i,j}$. In the approximate equality, the conditional independence of $c_{i,j}$ on $\boldsymbol{s}_j$ given $o_{i,j}$ (Eqn. 2) and the game outcome prediction (Eqn. 3) are used.

Now $c(\boldsymbol{s}_i, o_{i,j})$ can be efficiently learned as a standard churn prediction problem. The input features are merely an updated player state after knowing the matchmaking game outcome, i.e., $\boldsymbol{s}_i^{update} \leftarrow \boldsymbol{s}_i$ and $o_{i,j}$. If we decompose $\boldsymbol{s}_i = [\boldsymbol{o}_i^K, \hat{\boldsymbol{s}}_i]$, where $\boldsymbol{o}_i^K$ is a vector of the latest $K$ game outcomes (for example, $\boldsymbol{o}_i^K = LWLDL$ when $K = 5$), and $\hat{\boldsymbol{s}}_i$ represent the rest of features in $\boldsymbol{s}_i$. The state update can be done easily as:

$$\boldsymbol{s}_i^{update} \leftarrow \boldsymbol{s}_i \text{ and } o_{i,j} \qquad (7)$$

$$= [\boldsymbol{o}_i^K, \hat{\boldsymbol{s}}_i] \text{ and } o_{i,j} \qquad (8)$$

$$= [\boldsymbol{o}_i^{K+1}, \hat{\boldsymbol{s}}_i^{update}] \qquad (9)$$

We use $\hat{\boldsymbol{s}}_i^{update}$ to indicate that non-game-outcome features are also updated after a new match. For example, the total number of games played increments by one. We can adopt any churn prediction model based on the updated player state.

### 3.3 Finding the Optimal Pair Assignment

Given the predicted churn risks of each pair of players, i.e., the weight of every edge in $\mathcal{G}$, EOMM reduces to a minimum weight perfect matching (MWPM) problem. The goal is to find a pair assignment, $\mathcal{M}^*$, on a complete graph, $\mathcal{G}$, which has the minimal sum weights of edges.

For a graph with $N$ node, the brute-force way is to exhaustively compare all $\binom{N}{N/2}/2^{\frac{N}{2}}$ possible pair assignments and find the best one, but the time complexity is too high to be feasible in practical systems. Fortunately, many polynomial time algorithms exist for the MWPM problem. For example, several algorithms can solve the problem in the worst time complexity $O(N^3)$ [18, 26]. If engagement measurements are pure integers, there exists a slightly faster algorithm [19] with running time $O(N^{2\frac{3}{4}} \log K)$ where $K$ is the largest magnitude of an edge weight. There also exist greedy algorithms, such as [12] and [13], with faster running time to find suboptimal solutions. Moreover, MWPM can be solved in parallel as proposed by [35].

## 4. THEORETICAL FINDINGS

Besides generating optimal matchmaking assignments, EOMM provides a framework to conduct theoretical analysis on other matchmaking related problems. We use this framework to compare EOMM with other matchmaking strategies under different hypothetical situations to obtain many insights. Without loss of generality, we focus our discussion on 1-vs-1 games with possible game outcomes sampled from *Win*, *Lose* and *Draw*.

Using the same notation in Section 3, we investigate a pair of players $p_i, p_j \in \mathcal{P}, i \neq j$. When $c(\boldsymbol{s}_i, \boldsymbol{s}_j) = c(o_{i,j})$, i.e., a player's churn risk only depends on the game outcome of the upcoming match, regardless of all other states. This simplification for Eqn. 7, where $\boldsymbol{s}_i^{update}$ only considers $o_{i,j}$ but ignores $\boldsymbol{s}_i$, has interesting implications.

- If $c(Win) + c(Lose) > 2 \cdot c(Draw)$, i.e., the sum churn risk of two matched players in a tied game is lower than that in a non-tied game. Under this circumstance, the equal-skill based matchmaking is *equivalent* to EOMM, as both strive to form matches with *Draw* outcomes as many as possible. This explains the intuition and popularity behind equal-skill matchmaking. But we should be very aware of its conditional applicability, while EOMM is instead always optimal.

- If $c(Win) + c(Lose) < 2 \cdot c(Draw)$, equal-skill based matchmaking is actually *worst* among all matchmaking schemes, as its goal to create close matches contrarily minimizes the overall player engagement. Although this situation contradicts with the common intuition that fair matches are good, it is possible for a real game. Therefore validating the assumptions with real game data is critical before applying an equal-skill based matchmaking algorithm.

When $c(\boldsymbol{s}_i, \boldsymbol{s}_j) = c(\boldsymbol{s}_i)$, i.e., a player's churn risk is determined by his state before matchmaking, then it does not matter whom they will play. In this case, EOMM can do no better than a random matchmaking. Random matchmaking, from this perspective, is not as trivial as we thought. It is a relative safe and stable baseline choice in lack of prior information. While equal-skill based method can perform the worst under certain conditions, random matchmaking will never fall into the worst case.

The analysis above shows that the existing matchmaking methods, such as equal-skill based and random matching, arise within the EOMM framework on different conditions. Practitioners can safely apply EOMM while gathering more information about their game and players.

## 5. CASE STUDY

To test the proposed matchmaking framework, we ran simulation which is configured based on the real data from a popular PvP game made by Electronic Arts Inc. (EA). In the simulation, we compared different matchmaking methods applied to the same player population. In the end, EOMM retained significantly higher number of players than other matchmaking methods.

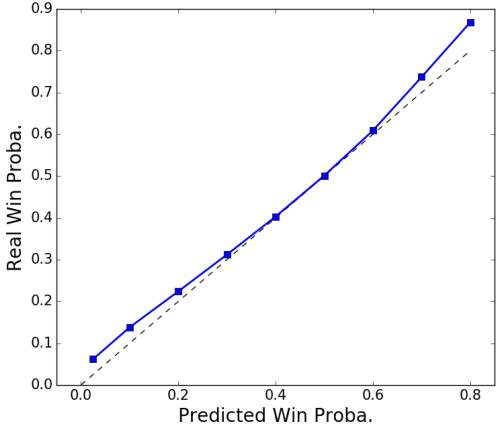

Figure 2: Predicted win probability vs. real win probability. Real win probability is the ratio of matches, with similar predicted winning probabilities, whose outcomes are real "Wins".

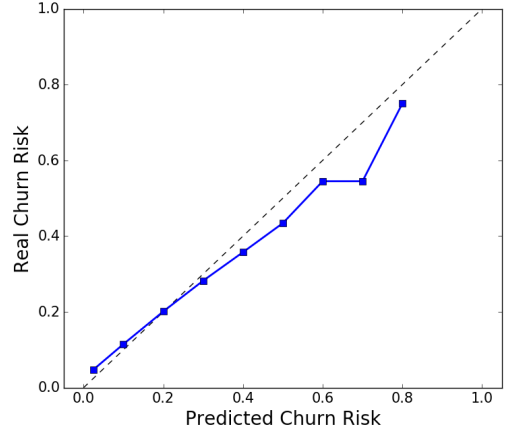

Figure 3: Predicted churn risk vs. real churn risk. Real churn risk is the ratio of matches, with similar predicted churn risks, which are indeed the last match before churn.

### 5.1 Data Collection

We collected 1-vs-1 matches from a popular game made by EA. There are three possible match outcomes, namely *Win*, *Lose* and *Draw*. In total, we collected 36.9 million matches played by 1.68 million unique players in the first half of 2016.

### 5.2 Preparation

To create a realistic environment for simulation, the following models and functions are needed. We compute them based on real game data.

**Player Skills** We need to establish a distribution of player skills for the population we simulate on. The distribution is learned from real game data. We sorted the collected real matches temporally and applied Glicko [20] to compute each player's final skill. For each player $i$, the skill vector is represented by mean $r_i$ and variance $RD_i$, i.e. $\boldsymbol{\mu}_i = (r_i, RD_i)$. In simulation, we assume that the game and player skills are stationary. The population's skill distribution is constant, where each player's skill does not change any more over time.

While Glicko scores can be used to estimate the winning probability of player $i$ over player $j$, $\Pr(i > j|\boldsymbol{\mu}_i, \boldsymbol{\mu}_j)$, they cannot provide the probability of draws. We defined a set of rules to allow the estimation of win/lose/draw probabilities from Glicko scores:

$$\Pr^*(i = j) = 20\% \qquad (10)$$

$$\Pr^*(i > j) = \frac{80\% \cdot \Pr(i > j|\boldsymbol{\mu}_i, \boldsymbol{\mu}_j)}{\Pr(i > j|\boldsymbol{\mu}_i, \boldsymbol{\mu}_j) + \Pr(j > i|\boldsymbol{\mu}_j, \boldsymbol{\mu}_i)} \qquad (11)$$

$$\Pr^*(i < j) = 1 - \Pr^*(i = j) - \Pr^*(i > j) \qquad (12)$$

Basically, the draw probability (Eqn. 10) is set to 20% regardless of skill gaps. This is based on our findings that 1) draw outcomes only have $-0.05$ correlation with the difference of skill means in the collected game data; 2) around 20% matches are draws regardless of skill gaps. The win/lose probabilities are normalized such that the probabilities of win, lose and draw sum up to 1. Figure 2 shows that the predicted win probabilities using Glicko scores based on our rules are well aligned with the real match outcomes.

**Churn Prediction Model** We trained a logistic regression model for predicting whether a player will be an eight-hour churner after a match. The input features describe the upcoming match and the player's 10 most recent matches. A player is labeled as an eight-hour churner if they do not play any 1-vs-1 match within the next eight hours after playing this match. As discussed in Section 3, the term of "churn" is used by convention. It represents "stopping playing" within a period of time, which is a metric of disengagement.

We use Eqn. 6 to estimate $c(\boldsymbol{s}_i, \boldsymbol{s}_j) + c(\boldsymbol{s}_j, \boldsymbol{s}_i)$. The model takes as input the player's state $\boldsymbol{s}_i$ before matchmaking along with the upcoming match outcome $o_{i,j}$.

Specifically, the input features consist of:

- *Each of the player's 10 most recent matches*: win/lose/draw status, time passage since the previous match, time passage to the upcoming match, and goal difference against his opponent

- *Upcoming match*: one-hot encoding of the upcoming match's outcome win/lose/draw

- *Other*: the number of 1-vs-1 matches played in the last eight hours, one day, one week and one month.

We use 5-fold cross validation and grid search to determine the proper $L_2$ regularization strength when training the model. The predicted probabilities are well aligned with the real churn probabilities, in particular when churn risk is less than 0.8, as shown in Figure 3. While the performance of the predictive model still has room to improve, the flexibility of EOMM allows one to easily refine or replace the model if better ones are found.

**Player States** In simulation, each player's state is sampled from a collection of states, which are established based on real players' states in the collected data. We first randomly sample a subset of matches. Both players' states in those matches are gathered to create this collection. A player state contains the needed features for churn prediction, as well as the player's skill score.

### 5.3 Simulation Procedure

In the simulation, we compared EOMM with three matchmaking mechanisms: *random matchmaking* (*RandomMM*), which randomly pairs available players in the waiting pool, *skill-based matchmaking* (*SkillMM*), which pairs every two consecutive players after sorting them by skills, and *worst matchmaking* (*WorstMM*), which does the opposite of EOMM by minimizing the objective function of EOMM. SkillMM always seeks "fair games". We added WorstMM as a validation.

All methods are applied on the same population (waiting pool), where the same player skill distribution, churn model and player state distribution as described in Section 5.2 are used. EOMM follows Eqn. 6 to estimate churn risk $c(s_i, s_j) + c(s_j, s_i)$. We used the perfect matching algorithm [18, 26] implemented by an open-source library [42].

For each matchmaking method $\mathcal{M}$, the procedure within each round of simulation is as follows:

1. Create a waiting pool of $P$ players, whose player states are sampled from the player state collection.

2. Use $\mathcal{M}$ to determine the pair assignment (matchmaking).

3. Simulate match outcomes according to the win/lose/draw probability predicted by the skill model

4. For each player, simulate if he will churn according to the predicted churn probability by churn model.

5. Record the number of retained players.

In experiments, we tested $P = 100, 200, 300, 400$ and $500$. For each setting of $P$, we repeated the simulation by 10,000 rounds of matchmaking. We compare different matchmaking methods by the average number of their retained players per round, i.e., the players who continue playing in the next eight hours. In order to test statistical significance, we conducted Welch's $t$-test between every pair of the matchmaking algorithms.

### 5.4 Results and Discussion

The results are shown in Table 2. All pairwise differences of retained players are statistically significant ($p$-value $< 0.01$) except EOMM vs. RandomMM (when $P = 100$) and SkillMM vs. RandomMM (when $P = 400$). In all other scenarios, EOMM outperforms the other three matchmaking methods. The results prove the applicability of EOMM to act as an engagement optimizer. When $P = 100$, EOMM does not retain a significantly higher number of players than RandomMM, and even retains fewer players than SkillMM. It is possibly because that when $P$ is small, the randomness has higher impact, and also, the room for arranging opponents is smaller. More rounds of simulations might be needed to show significance in this case.

The improvement of EOMM over SkillMM, the most common matchmaking method, in terms of the average number of retained players are 0.3%, 0.9%, 1.1%, and 0.6% when $P = 200, 300, 400$ and 500 respectively. On average EOMM retains 0.7% more player compared with SkillMM after one round of matchmaking. Notably the benefit of retention will accumulate over time in a constant population. For players who play 20 rounds of matchmaking games within eight hours, there will be 15% more players retained ($1.007^{20} \approx 1.15$) by EOMM over those by SkillMM. The more rounds of matchmaking are conducted, the more significant is the accumulative advantage of EOMM in engagement.

Table 2: Average number of retained players per round of matchmaking simulation. 10,000 rounds of matchmaking were simulated.

| *Method* | P=100 | P=200 | P=300 | P=400 | P=500 |
|---|---|---|---|---|---|
| WorstMM | 51.50 | 103.39 | 154.57 | 206.65 | 258.21 |
| SkillMM | 52.52 | 103.96 | 156.05 | 207.43 | 259.24 |
| RandomMM | 51.81 | 103.97 | 156.09 | 207.09 | 259.65 |
| EOMM | 51.90 | 104.24 | 157.50 | 209.37 | 261.19 |

We did not find a consistent climb in retention boost as $P$ increased. This may suggest that when the player pool reaches certain size, the choices of opponents are enough to rescue those players on the edge of churn. Beyond this size, a larger player pool may not bring in significantly extra benefits in engagement maximization.

As a validation, WorstMM consistently retains the fewest players in the pools of all sizes. This result verifies the optimum of EOMM from the opposite side. It is also interesting to note that SkillMM does not consistently outperform RandomMM, which is aligned with our discussion in the theoretical findings in Section 4, that is, balanced matches are not always optimal for engagement.

## 6. CONCLUSION

The paper presents a novel framework to achieve engagement optimized matchmaking (EOMM). It formulates matchmaking as a problem of maximizing the player engagement, and solves the optimization efficiently. EOMM employs three components, a skill model, an engagement predictive model and a minimum weight perfect matching algorithm, each of which can be tailored flexibly for specific applications. We ran simulations whose configurations were based on real data from an online PvP game. The results show that EOMM significantly outperforms all other methods in the number of retained players. EOMM also provides a theoretical framework to analyze various matchmaking algorithms.

EOMM provides a measurable and flexible matchmaking framework. It has well-defined quantitative objectives that can be monitored, evaluated and improved. Within the EOMM framework, the core building components, skill model, churn model and graph pairing model, are uncoupled so that they can be tuned and replaced independently. Moreover, we can even change the objective function to other core game metrics of interest, such as play time, retention, or spending. EOMM allows one to easily plug in different types of predictive models to achieve the optimization.

So far we have discussed EOMM in 1-vs-1 game scenarios. This framework also applies to PvP games that involve teams of players, where every component needs to be extended with additional care. The skill model can be simply applied to a team by adding up skills for all team members [23]. For churn prediction, we can use the same idea that one player's churn risk is conditionally independent with other players' states given that their influence on the player's own state, such as the game outcome, is known. Last, the minimum weight perfect matching algorithms for pairs are no longer applicable. Instead of a pair assignment, we seek a *grouping assignment* on a complete graph. A related area to investigate is perfect matching in hypergraphs [5], where an edge can connect more than two vertices. Furthermore, EOMM is not even limited to games. In broad applications, such as friend connection in a social network and 1-on-1 tutoring in online education, EOMM's formulation and optimization techniques still apply.

In the future, we expect EOMM equipped with more advanced models, such as skill model and churn model, can have higher optimal bound. We will explore EOMM performance in more realistic situations, where practical restrictions are applied, such as network connectivity, regions and friend/black lists. More restrictions would result in fewer edges in the constructed graph of EOMM. Last, we will explore EOMM in multi-player games with more than two players involved and efficient algorithms analogous to perfect matching algorithms within hypergraphs.

## 7. ACKNOWLEDGMENTS

This work is partially supported by NSF Career #1453800.